\documentclass[a4paper,10pt]{article}
\usepackage{graphicx} 
\usepackage{graphics}

\usepackage{amsmath}   
\usepackage{amsfonts}
\usepackage{amssymb}

\begin{document}
%
%
\def\ov{\over}
\def\l{\left}
\def\r{\right}
\def\be{\begin{equation}}
\def\ee{\end{equation}}
%

\setlength{\oddsidemargin} {1cm}
\setlength{\textwidth}{18cm}
\setlength{\textheight}{23cm}
\title{Electromagnetic Klein-Gordon and Dirac equations in scale relativity}
\author{{\bf Marie-No\"elle C\'el\'erier$^1$ and Laurent Nottale$^2$} 
\\
Laboratoire Univers et Th\'eories (LUTH), \\
Observatoire de Paris, CNRS, Universit\'e Denis Diderot, \\
5 place Jules Janssen, 92190 Meudon, France \\
e-mail: $^1$ marie-noelle.celerier@obspm.fr \\
$^2$ laurent.nottale@obspm.fr}


\maketitle

\begin{abstract}
We present a new step in the foundation of quantum field theory with the tools of scale relativity. Previously, quantum motion equations (Schr\"odinger, Klein-Gordon, Dirac, Pauli) have been derived as geodesic equations written with a quantum-covariant derivative operator. Then, the nature of gauge transformations, of gauge fields and of conserved charges have been given a geometric meaning in terms of a scale-covariant derivative tool. Finally, the electromagnetic Klein-Gordon equation has been recovered with a covariant derivative constructed by combining the quantum-covariant velocity operator and the scale-covariant derivative. We show here that if one tries to derive the electromagnetic Dirac equation from the Klein-Gordon one as for the free particle motion, i.e. as a square root of the time part of the Klein-Gordon operator, one obtains an additional term which is the relativistic analog of the spin-magnetic field coupling term of the Pauli equation. However, if one first applies the quantum covariance, then implements the scale covariance through the scale-covariant derivative, one obtains the electromagnetic Dirac equation in its usual form. This method can also be applied successfully to the derivation of the electromagnetic Klein-Gordon equation. This suggests it rests on more profound roots of the theory, since it encompasses naturally the spin-charge coupling.
\end{abstract}


\section{\bf Introduction}
\label{intro}

The theory of scale relativity generalizes to scale transformations the principle of relativity, which has been applied by Einstein to motion laws. It is based on giving up the manifold differentiability assumption, which is a key hypothesis in general relativity. In the new theory, coordinate transformations are continuous and can be differentiable (including  therefore general relativity) or nondifferentiable. This implies several consequences \cite{LN93}, leading to successive steps of the theory construction. We consider here the developments applying only to particle physics and field theory in the relativistic quantum domain.

Continuity and non-differentiability imply space-time fractality \cite{LN93,BC00}, which we define as a scale dependence of the reference frames. Therefore, the physical quantities, and in particular the coordinates themselves, become functions of new scale variables, added to the usual position, orientation and motion variables defining the reference frames. In a theoretical physics description, these new variables are identified with the differential elements \cite{CN04}; hence, we write $x^{\mu} = x^{\mu} ({\rm d}x^{\mu})$ and a physical quantity $f$ is represented by a fractal function $f[x^{\mu} ({\rm d}x^{\mu}),  {\rm d}x^{\mu}]$.

The scale variables are defined only in a relative way; namely, only their ratio has a physical meaning. This leads to extending to scales the principle of relativity and to including in the possible changes of the reference frames those implied by transformations of these scale variables. As a first step of the development of the theory and for simplification purpose, we have proposed that at each point in space-time is `attached' an intrinsic `scale-space' in which the scale transformations are defined. This `scale-space' is a mathematical analog of a fiber space, but the transformations described in its formalism are physical transformations occurring in the actual space-time. Though nondifferentiability manifests itself everywhere only at the limit where the scale variables vanish ($ {\rm d}x^{\mu} \to 0$), the use of differential equations of scale (which describe the way they tend to zero, i.e., what happens in a differential scale transformation $ \ln | {\rm d}x^{\mu}| \to \ln | {\rm d}x^{\mu}| + {\rm d} \ln | {\rm d}x^{\mu}|$), constrained by the relativity principle, is made possible and various scale laws derived from these differential equations have been studied \cite{LN92,LN96,LN97}.

In the following, we restrict ourselves to the simplest case of Galilean-like scale transformations characterized by a constant fractal dimension, $D_F = 2$. This choice relies on a feature exhibited by the typical paths of quantum particles which contribute mainly to the path integral: these are nondifferentiable and fractals of dimension $D_{F}=2$ \cite{RF48,FH65}. Actually, the case of a variable fractal dimension has also been considered and thoroughly studied \cite{LN92,LN96},
but $D_{F}=2$ can be regarded as a good approximation for the whole scale domain under consideration here.

The transition from the classical to the quantum domain emerges when we consider the simplest scale differential equation which is a first order renormalization-group-like equation and when we Taylor-expand its Callan-Symanzik-like function at first order in powers of the coordinates or of the components of the 4-velocity \cite{LN93,CN04,NCL06}.

In a fractal space-time, the geodesic equations are also scale-dependent and the number of geodesics that relate any two events (or start from any event) tends to infinity \cite{LN93,NC07}. We adopt therefore a generalized statistical fluid-like description where the standard velocity $V^{\mu}(s)$ is replaced by a scale-dependent, fractal velocity field $V^{\mu}[x^{\mu}(s,{\rm d}s),s,{\rm d}s]$, where ${\rm d}s$ denotes the intervals of the invariant length, i.e. the proper time $s$, on the geodesics.

It has been shown that, for relativistic motion, the transition from the classical to the quantum domain occurs, in the rest frame, around the Compton scale $\lambda_c$ of the `particle', the quantum domain being defined by ${\rm d} s < \lambda_c$ \cite{CN04,LN07}.

The next step consists in writing the geodesic equation. We make the conjecture that the `internal' properties of quantum `particles' are given by the geometrical properties of the geodesic bundle corresponding to their state, according to the various conservative quantities (prime integrals) which define them. Any measurement performed on the `particle' is interpreted as a selection of the geodesic bundle linked to the interaction with the measurement apparatus (which depends on its resolution) and/or to the information known about it (for example, the which-way information in a two-slit experiment \cite{LN96,NC07}).

Another consequence of nondifferentiability is the breaking of the invariance by reflexion of the differential element ${\rm d}s$. Indeed, in terms of fractal functions $f(s,{\rm d}s)$, two generalized derivatives are defined instead of one and a complex quantum-covariant derivative is constructed which includes the various effects of nondifferentiability and fractality described above \cite{CN04,LN96}. Using a generalized equivalence principle, formally identical to a strong covariance principle, we can write a geodesic equation in terms of this covariant derivative and obtain the Klein-Gordon equation for a free particle \cite{CN04,LN96}.

Then we introduce further symmetry breakings, that of the invariance by reflexion of the differential element ${\rm d}x^{\mu}$ and of the parity and time reflexion symmetries. These allow us to obtain the free Dirac equation of standard quantum mechanics \cite{CN04}. At the nonrelativistic limit the Pauli equation is also recovered with the proper value of the electron magnetic moment \cite{CN06}.

The further step is to consider the coupling between motion in the fractal space-time and transformations of the scale variables. According to the scale relativity principle, the scale-space is fundamentally nonabsolute, i.e. the scale of a structure (internal to the fractal geodesics which are identified with a `particle') is expected to change during a displacement in space-time. In other words, the scale variables are now interpreted as explicit functions of the coordinates, such as the invariant ${\rm d}s = {\rm d}s(t,x,y,z)$.

In previous works, the nature of gauge transformations, of gauge fields and of the conserved charges in both the Abelian \cite{LN96,LN94,LN03} and the non-Abelian \cite{NCL06} cases have been given a physical meaning as a consequence of this coupling. Then the electromagnetic Klein-Gordon equation has been recovered from a geodesic equation in a fractal space-time where the covariant derivative is constructed through the combination of a complex quantum-covariant velocity operator and the scale-covariant derivative \cite{LN07}.

The present article describes a new step in the foundation of field theory from the scale relativity first principles. To compare the electromagnetic Dirac equation with the electromagnetic Klein-Gordon equation in a manner analogous to the scheme adopted in \cite{CN04} for the free particle motion, we first establish a biquaternionic electromagnetic Klein-Gordon-like equation. Then, we compare it with the equation obtained when squaring the time part of the biquaternionic electromagnetic Dirac operator and applying it to the wave function. We obtain thus an additional term which is the relativistic analog of the spin-magnetic field coupling term encountered in the Pauli equation. This means that, if we merely combine the two covariances as done previously for the Klein-Gordon equation when a nonzero spin particle is involved, an additional coupling term appears in the Dirac equation. Such a nonlinearity of the covariant derivatives is not unexpected: indeed, their effect is clearly distinct only in the spinless case (one giving rise to the quantum effects and the other to the gauge charges and fields). But adding in the quantum-covariant derivative the effects of the reflexion symmetry breaking  ${\rm d}x^{\mu} \leftrightarrow \, -{\rm d}x^{\mu}$ \cite{CN04,CN06} leads to the emergence of spin which, through its connection with the magnetic moment, becomes coupled to the magnetic field, thus owning properties of an effective charge, so that the action of the two covariant derivatives is no longer independent.

To get rid of this drawback, we combine otherwise the quantum and scale covariances by first applying the quantum covariance such as to obtain the free Dirac equation as in \cite{CN04,CN03}, and then by implementing the scale covariance through the QED covariant derivative previously derived with the scale relativity tools. Thus we obtain the electromagnetic Dirac equation in its usual form.

This paper is organized as follows. In Sec.~\ref{esr}, we give a reminder of the construction of electrodynamics and of the QED covariant derivative in the framework of scale relativity. In Sec.~\ref{cekgge}, a short summary of the derivation of the standard electromagnetic Klein-Gordon equation with a complex wave function is provided. In Sec.~\ref{ede}, we first construct a biquaternionic electromagnetic Klein-Gordon-like equation and compare it to the Dirac equation for an electron/positron immersed in an electromagnetic field. This yields a spin-charge coupling term which encompasses the spinorial nature of the electron (positron). Then, we recover the electromagnetic Dirac equation in its standard form by combining in a different way quantum and scale covariances. Section~\ref{dc} is devoted to our conclusions.


\section{\bf Electrodynamics in Scale Relativity. A Reminder}
\label{esr}


\subsection{\bf \it Electromagnetic field and electric charges}

In the Abelian case corresponding to electromagnetism, the set of scale variables
has only one element, $\varrho=\lambda/\varepsilon$, where $\varepsilon$ is a given scale interval and $\lambda$ is a reference scale needed to make $\varrho$ nondimensional and to implement the scale relativity principle. This implies to limiting ourselves to the study of global scale transformations (contractions/dilations) in `scale-space' \cite{LN96,LN94,LN03}.

We have seen in Sec.~\ref{intro} that, according to the scale relativity principle, the scale of a structure internal to the fractal geodesics is expected to change during a displacement in space-time. Hence, scale variables are explicit functions of the coordinates, i.e. ${\rm d}s = {\rm d}s(t,x,y,z)$. We will see below that this applies in the domain where ${\rm d} s < \lambda_c$, i.e. in the same application domain as that of quantum particle relativistic motion.

Therefore, we expect, in a displacement, the appearance of a scale interval change due to the fractal geometry, which can be written as
\be
\delta \varepsilon =-\frac{1}{q}\;A_{\mu }\;\varepsilon \;{\rm d}x^{\mu },
\label{eq1}
\ee
i.e. in terms of the scale ratio:
\be
\delta \ln \varrho =\frac{1}{q}\;A_{\mu }\;{\rm d}x^{\mu }.
\label{eq2}
\ee

Here, $q$ is the `active' electric charge \cite{NCL06,LN03}. Defining $\chi =q \ln \varrho$, this leads to the appearance of a dilation field, issuing from the construction of a scale-covariant derivative,
\be
{\rm D} \chi ={\rm d}\chi -\delta \chi ={\rm d}\chi -A_{\mu }{\rm d}x^{\mu }.
\label{eq3}
\ee

Finally, we obtain the partial derivative as the sum of the inertial and geometric terms, i.e.
\be
\partial _{\mu }\chi = {\rm D}_{\mu }\chi +A_{\mu }.
\label{eq4}
\ee

In the framework of any space-time theory based on a relativity principle, the variation of the action $S$ of a particle is given directly by the space-time invariant ${\rm d}s$, i.e. $\delta \int {\rm d}S=0$ becomes a geodesic (Fermat) principle, $\delta \int {\rm d}s=0$ \cite{NCL06,LL72}. But here the fractality of the geodesics, with which the particle wave field is identified, means that their proper length is a function of the scale variable, so that $S=S(\chi )$. Therefore, the differential of the action reads
\be
{\rm d}S=\frac{\partial S}{\partial \chi }\,{\rm d}\chi =\frac{\partial S}{\partial \chi
}\,({\rm D}\chi +A_{\mu }{\rm d}x^{\mu }),
\label{eq5}
\ee
so that we obtain
\be
\partial _{\mu }S={\rm D}_{\mu }S+\frac{\partial S}{\partial \chi }\;A_{\mu }.
\label{eq6}
\ee

This result provides us with a definition for the `passive' charge, $e$ (on which the electromagnetic field acts), given by $e/c = - \partial S / \partial \chi$ \cite{LN96,NCL06,LN94}. A scale-relativistic equivalence principle, set to account for the action-reaction principle in Coulomb's law, implies $e = q$ and the above definition becomes a definition for the fine structure constant $\alpha = e^2/\hbar c = -\partial (S/\hbar)/\partial \ln \varrho$.

We have therefore recovered from the first principles of the theory the form of the action in standard electromagnetism as
\be
{\rm d}S=-mc\,{\rm d}s-\frac{e}{c}\,A_{\mu }\,{\rm d}x^{\mu }.
\label{eq7}
\ee

\subsection{\bf \it Quantum electrodynamics}

However, this action is not yet complete, since it should also contain the field term which we consider now.

As in any relativity theory, the action is equivalent to the space-time invariant, ${\rm d}S = - mc \; {\rm d}s$. The total elementary length, $ {\rm d}s_{tot}$, i.e. the proper time in fractal space-time, reads here
\be
{\rm d}s_{tot} = (h_{\mu \nu} \; {\rm d}x^{\mu}{\rm d}x^{\nu})^{1/2} + \frac{e}{mc^2} A_{\mu} {\rm d}x^{\mu},
\label{eq8}
\ee
where $h_{\mu \nu}$ is the Minkowski metric tensor. It is worth recalling the meaning of this expression in the scale-relativistic framework. The length of a fractal path can change, first because its extremity has moved, which is interpreted in terms of the particle motion (this is expressed by the first term on the r.h.s. of Eq.~(\ref{eq8}), which is the same for fractal and non-fractal geometry), but also because of internal dilations/contractions (this is expressed by the second term). In extreme cases, there can be a purely internal length increase with no space displacement counterpart which might be interpreted as a potential energy, or a purely internal length decrease manifesting itself in terms of motion \cite{LN07}.

Then, we use the action-geodesic principle by postulating the following generalized geodesic equation applied to the full invariant proper length
\be
\delta \int {\rm d} s_{tot} = 0.
\label{eq9}
\ee

Although we have recovered the standard variational principle of electromagnetism, we will nevertheless develop it hereafter so as to be allowed to give a new geometric interpretation of the electromagnetic field and to define the covariant derivative of a vector which will be needed subsequently to derive the electromagnetic Klein-Gordon equation in this framework.

We write the invariant length/proper time variation as
\be
\delta s_{tot} = \int \left(\frac{{\rm d}x_{\mu}{\rm d}\delta x^{\mu}}{{\rm d} s} + \frac{e}{mc^2} A_{\mu} {\rm d}\delta x^{\mu} + \frac{e}{mc^2} \delta A_{\nu} {\rm d}x^{\nu}\right) = 0.
\label{eq10}
\ee

After integration by parts between two fixed values of the coordinates taken as bounds, Eq.~(\ref{eq10}) becomes
\be
\int \left( {\rm d}u_{\mu} \delta x^{\mu} + \frac{e}{mc^2} {\rm d} A_{\mu} \delta x^{\mu} - \frac{e}{mc^2} \delta A_{\nu}{\rm d}x^{\nu}\right) = 0.
\label{eq11}
\ee

Since $\delta A_{\nu} = (\partial A_{\nu}/\partial x^{\mu})\delta x^{\mu}$ and ${\rm d} A_{\mu} = (\partial A_{\mu}/ \partial x^{\nu}) {\rm d}x^{\nu}$, we obtain
\be
\int \left\{ \frac{{\rm d}u_{\mu}}{{\rm d} s} - \frac{e}{mc^2} \left(\frac{\partial A_{\nu}}{\partial x^{\mu}} - \frac{\partial A_{\mu}}{\partial x^{\nu}}\right) u^{\nu} \right\} \delta x^{\mu} {\rm d} s = 0.
\label{eq12}
\ee

All the terms additional to the inertial ones have their geometric origin in scale transformations. We are therefore led to define the scale-covariant differential of the velocity $u_{\nu}$ as
\be
{\rm D} u_{\nu} = {\rm d} u_{\nu} - \frac{e}{mc^2} F_{\nu \mu} {\rm d}x^{\mu},
\label{eq13}
\ee
where the `connection', $F_{\nu \mu} = \partial A_{\mu}/ \partial x^{\nu} - \partial A_{\nu}/ \partial x^{\mu}$, can be identified with the electromagnetic tensor.

We are now able to define the scale-covariant partial derivative of a scale-variable-dependent vector, $B_{\nu}$, as \cite{LN07}
\be
{\rm D}_{\mu} B_{\nu} = \partial_{\mu} B_{\nu} + \frac{e}{mc^2} F_{\mu \nu}.
\label{eq14}
\ee

Applying a generalized strong covariance principle, which extends the covariance principle of general relativity, the motion equation of electrodynamics is established as a geodesic equation which keeps the form of a free Galilean equation of motion in terms of the scale-covariant derivative \cite{LN07}
\be
\frac{{\rm D} u_{\nu}}{{\rm d}s} = u^{\mu} {\rm D}_{\mu} u_{\nu} = 0.
\label{eq15}
\ee

Expanding the expression of the covariant derivative on the l.h.s. of Eq.~(\ref{eq15}), we obtain
\be
\frac{{\rm d}u_{\nu}}{{\rm d}s} - \frac{e}{mc^2} F_{\nu \mu} u^{\mu} = 0,
\label{eq16}
\ee
which is nothing else that the Lorentz motion equation in electromagnetism.

Let us now come back to the issue of characterizing the transition to the gauge theory domain where a small increment of the length invariant ${\rm d} s$ becomes dependent on the space-time coordinates. Note that the scale-relativistic approach shares some features with the Weyl-Dirac theory of electromagnetism \cite{HW18,PD73}. However, in this last theory, the variation of ${\rm d} s$ is postulated to exist at all scales, contradicting the observed invariance of the electron mass, and thus of its Compton length. In scale relativity, the effects of the coordinate dependence of scale variables are observable only below the fractal/nonfractal transition, which is identified in rest frame with the particle Einstein time scale, $\tau_E = \hbar /mc^2$. This time scale corresponds, up to the fundamental constant $c$, to the Compton scale $\lambda_c = \hbar /mc$, which is the quantum/classical transition scale for relativistic motion. This means that for scales smaller than $\lambda_c$ the fractal nondifferentiable fluctuations dominate the classical behavior; hence the influence of fractality both on motion and on scale-motion coupling.

\subsection{\bf \it QED covariant derivative}
\label{qcd}

We consider a generalized action which depends on both motion and scale variables. In the scale-relativistic approach to quantum theory, the 4-velocity ${\mathcal V}^{\mu }$, which describes a scalar particle, is complex, so that its action is also complex and can be written $S=S(x^{\mu },{\mathcal V}^{\mu },\chi )$. The wave function is defined in terms of this action as $\psi =\exp \left( i{S}/{\hbar }\right)$ \cite{LN96}.

Therefore, Eq.~(\ref{eq7}) takes the form,
\be
{\rm d} S= - mc \; {\mathcal V}_{\mu } \; {\rm d} x^{\mu } - \frac{e}{c} A_{\mu }{\rm d} x^{\mu } = -i \hbar \; {\rm d} \ln \psi.
\label{eq17}
\ee

We thus obtain a new relation between the complex velocity and the wave function:
\be
{\mathcal V}_{\mu } = i \lambda_c \; {\rm D}_{\mu}(\ln \psi) = i \lambda_c \; \partial_{\mu}(\ln \psi) - \frac{e}{mc^2}A_{\mu}.
\label{eq18}
\ee

We recover here and give a geometric foundation to the standard QED covariant derivative,
\be
{\rm D}_{\mu} = \partial_{\mu} + i \frac{e}{\hbar c}A_{\mu},
\label{eq19}
\ee
 as being the scale-covariant derivative now acting on the wave function.


\section{\bf Complex Electromagnetic Klein-Gordon Equation as a Geodesic Equation}
\label{cekgge}


Now, we proceed with a generalization of the scale relativity approach to QED. Here, both the quantum and electromagnetic properties are expected to emerge from the fractal geometry of space-time. We therefore combine the quantum-covariant derivative which describes the effects induced by fractality on motion and the scale-covariant derivative which describes scale-motion coupling, i.e. nonlinear effects of the coordinate dependence of scale variables, to obtain a single covariant tool. Then, as for the free motion case \cite{CN04,LN96}, we write the Klein-Gordon equation for a particle submitted to an external electromagnetic field as a free geodesic equation exhibiting the inertial Galilean form ${\rm D}V/{\rm d} s = 0$, where the quantum behavior and the field are both generated by the doubly covariant derivative ${\rm D}$.

As regards quantum covariance, it is implemented in the simplest motion-relativistic case (corresponding to a single symmetry breaking ${\rm d}s \leftrightarrow - \; {\rm d}s$ in the scale relativity formalism) by the use of a covariant complex velocity operator: \cite{LN07}
\be
\widehat{{\mathcal V}^{\mu }} = {\mathcal V}^{\mu } + i(\lambda_c/2) \partial^{\mu}.
\label{eq20}
\ee

In Sec.~\ref{esr}, we recalled how the electromagnetic field can be constructed with the use of a scale-covariant derivative ${\rm D}_{\mu}$ which implements dilation/contraction transformations in the fractal space-time. We combine now the two tools and we define a doubly (quantum and scale) covariant derivative as
\be
\frac{\widehat{\rm D}}{{\rm d} s} = \widehat{{\mathcal V}^{\mu }} {\rm D}_{\mu},
\label{eq21}
\ee
which we use to write the following inertial-like, strongly covariant geodesic equation
\be
\frac{\widehat{\rm D}}{{\rm d} s} {\mathcal V}_{\nu} = 0.
\label{eq22}
\ee

This very simple, free-form equation gives, after integration, the Klein-Gordon equation in the presence of an external electromagnetic field. The first step amounts to obtain a quantum-covariant analog of the Lorentz equation. Successively developing the covariant derivatives in Eq.~(\ref{eq22}), we obtain
\be
\frac{\widehat{\rm D}}{{\rm d} s} {\mathcal V}_{\nu} = \widehat{{\mathcal V}^{\mu }} {\rm D}_{\mu} {\mathcal V}_{\nu} = \widehat{{\mathcal V}^{\mu }}\left(\partial_{\mu} {\mathcal V}_{\nu } +  \frac{e}{mc^2} F_{\mu \nu}\right) = 0.
\label{eq23}
\ee

Since $\widehat{\rm d}/{\rm d} s = \widehat{{\mathcal V}^{\mu }} \partial_{\mu}$ \cite{LN07}, we can write $\widehat{{\mathcal V}^{\mu }} \partial_{\mu} {\mathcal V}_{\nu } = \widehat{\rm d} {\mathcal V}_{\nu }/ {\rm d} s$. Thus, Eq.~(\ref{eq23}) becomes
\be
mc \frac{\widehat{\rm d}}{{\rm d} s} {\mathcal V}_{\nu} = \frac{e}{c} \widehat{{\mathcal V}^{\mu }} F_{\nu \mu},
\label{eq24}
\ee
which exhibits the exact form of the standard Lorentz equation of dynamics but written with quantum-covariant derivative and velocity.

To proceed with the derivation of the Klein-Gordon equation, we first note that, thanks to Eq.~(\ref{eq18}),
\be
\partial_{\mu} {\mathcal V}_{\nu} - \partial_{\nu} {\mathcal V}_{\mu} = - \frac{e}{mc^2}(\partial_{\mu} A_{\nu} - \partial_{\nu} A_{\mu}) = - \frac{e}{mc^2} F_{\mu \nu}.
\label{eq25}
\ee

We can therefore replace, in Eq.~(\ref{eq23}), $\partial_{\mu} {\mathcal V}_{\nu} + (e/mc^2) F_{\mu \nu}$ by $\partial_{\nu} {\mathcal V}_{\mu}$. This gives
\be
\widehat{{\mathcal V}^{\mu }} \partial_{\nu} {\mathcal V}_{\mu } = 0.
\label{eq26}
\ee

Now, we first develop the quantum-covariant velocity operator in Eq.~(\ref{eq26}), and then we replace ${\mathcal V}^{\mu }$, respectively ${\mathcal V}_{\mu }$, by their scale-covariant form given by Eq.~(\ref{eq18}) and we obtain
\be
\left(i \lambda_c \; \partial^{\mu}(\ln \psi) -  \frac{e}{mc^2} A^{\mu} + i \frac{\lambda_c}{2} \, \partial^{\mu} \right) \partial_{\nu} \left(i \lambda_c \; \partial_{\mu}(\ln \psi) -  \frac{e}{mc^2} A_{\mu}\right) = 0.
\label{eq27}
\ee

Replacing the Compton length $\lambda_c$ by its expression $\hbar / mc$ and using the identity
\be
\left(\partial^{\mu}(\ln \psi) + \frac{1}{2}\,  \partial^{\mu}\right) \partial_{\nu}\partial_{\mu} (\ln \psi) = \frac{1}{2} \partial_{\nu} \left(\frac{\partial^{\mu} \partial_{\mu} \psi}{\psi} \right),
\label{eq28}
\ee
we obtain, after some rearrangements,
\be
\partial_{\nu}\left[- \hbar^2 \left(\frac{\partial^{\mu} \partial_{\mu} \psi}{\psi} \right) - 2i \hbar \, \frac{e}{c} A_{\mu}\frac{\partial^{\mu} \psi}{\psi} + \frac{e^2}{c^2} A^{\mu}A_{\mu} - i \hbar \, \frac{e}{c} \, \partial^{\mu}A_{\mu}\right] = 0.
\label{eq29}
\ee

This equation can be integrated and if we choose the integration constant to be $m^2 c^2$, this gives, after some rearrangements,
\be
- \hbar^2 \partial^{\mu} \partial_{\mu} \psi - i \hbar \, \frac{e}{c} \, \partial_{\mu} (A^{\mu} \psi) - i \hbar \, \frac{e}{c} A_{\mu} \partial^{\mu} \psi + \frac{e^2}{c^2} A^{\mu}A_{\mu} \psi = m^2 c^2 \psi,
\label{eq30}
\ee
which is actually the electromagnetic Klein-Gordon equation for a complex wave function:
\be
\left(i \hbar \, \partial_{\mu} - \frac{e}{c} A_{\mu}\right)\left(i \hbar \, \partial^{\mu} - \frac{e}{c} A^{\mu}\right) \psi = m^2 c^2 \psi.
\label{eq31}
\ee


\section{\bf Electromagnetic Dirac Equation}
\label{ede}


\subsection{\bf \it Spin-charge coupling}
\label{scc}

It has been known for a long time that the free Dirac equation proceeds from the free Klein-Gordon equation when written in a quaternionic form \cite{CL29,AC37}. In \cite{CN04,CN03} we have proposed to introduce naturally, as a consequence of the nondifferentiable geometry, a biquaternionic covariant derivative operator, leading to the definition of biquaternionic velocity and wave function, which we have used to derive a free Klein-Gordon-like equation in a biquaternionic form. Then, the free Dirac equation in a biquaternionic form, i.e. a bispinor form, follows. We show in the present section that if this method is applied when dealing with an exterior magnetic field, a new term corresponding to the coupling between the spin and the magnetic moment of the electron shows up. 

We use the quaternionic formalism, introduced by Hamilton \cite{WH66} and further developed by Conway \cite{AC37,AC45} (see also Synge \cite{JS72}). Our choice of the metric signature is (+ -- -- --).

\subsubsection{\it Biquaternionic electromagnetic Klein-Gordon-like equation}
\label{bqekg}

Considering the full consequence of nondifferentiability at the deepest level involves the subsequent breaking of the symmetries \cite{CN04,CN03}: \\
\begin{center}
${\rm d}s\leftrightarrow - \, {\rm d}s$ \\
${\rm d}x^{\mu}\leftrightarrow - \, {\rm d}x^{\mu}$ \\
$x^{\mu}\leftrightarrow - \, x^{\mu}$.
\end{center}

The 4-velocity becomes biquaternionic so that its action is also biquaternionic and reads ${\cal S} = {\cal S}(x^{\mu}, {\cal V}^{\mu}, \chi)$. We define the wave function as a re-expression of this action, i.e.
\be
\psi^{-1} \partial_{\mu} \psi = \frac{i}{\hbar} \partial_{\mu} {\mathcal S}.
\label{eq32}
\ee

Since the action must verify
\be
\delta {\mathcal S} = \partial_{\mu} {\mathcal S} \, \delta x^{\mu} = -mc \, {\mathcal V}_{\mu} \, \delta x^{\mu},
\label{eq33}
\ee
the 4-velocity reads
\be
{\mathcal V}_{\mu} = i \lambda_c \, \psi^{-1} \partial_{\mu} \psi.
\label{eq34}
\ee

The quantum-covariant derivative operator for relativistic motion is \cite{CN04,CN03}
\be
\frac{\widehat{\rm d}}{{\rm d} s} = {\mathcal V}^{\mu} \partial_{\mu} + i \frac{\lambda_c}{2} \, \partial^{\mu} \partial_{\mu}.
\label{eq35}
\ee
Therefore, quantum-covariance can still be implemented by means of Eq.~(\ref{eq20}), but now with a bi-quaternionic 4-velocity.

As in Sec.~\ref{cekgge}, we combine the scale-covariant derivative, as given by Eq.~(\ref{eq14}), with the quantum-covariant velocity, as given by Eq.~(\ref{eq20}), under the form of the geodesic equation (\ref{eq22}), which gives Eq.~(\ref{eq23}),
but here with a biquaternionic 4-velocity.

The form of the action for a particle in an electromagnetic field as given by Eq.~(\ref{eq7}) allows us to write
\be
\partial_{\mu} {\mathcal S} = - mc {\mathcal V}_{\mu} -\frac{e}{c} A_{\mu} = - i \hbar \psi^{-1} \partial_{\mu} \psi,
\label{eq36}
\ee
which gives
\be
{\mathcal V}_{\mu} = i \lambda_c \psi^{-1} \partial_{\mu} \psi - \frac{e}{mc^2} A_{\mu}.
\label{eq37}
\ee

We insert this expression for ${\mathcal V}_{\mu}$ in the biquaternionic version of Eq.~(\ref{eq23}) and obtain, after some rearrangements,
\be
\left(i \lambda_c  \psi^{-1} \partial^{\mu} \psi - \frac{e}{mc^2} A^{\mu} + i \frac{\lambda_c}{2} \partial^{\mu}\right)\left[i \lambda_c (\partial_{\mu} \psi^{-1} \partial_{\nu} \psi + \psi^{-1} \partial_{\mu} \partial_{\nu}\psi) - \frac{e}{mc^2} \partial_{\nu} A_{\mu}\right] = 0.
\label{eq38}
\ee

The definition of the inverse of a (bi)quaternion,
\be
\psi \psi^{-1} =  \psi^{-1} \psi = 1,
\label{eq39}
\ee
implies that $\psi$ and $\psi^{-1}$ commute. But this is not necessarily the case for $\psi$ and $\partial _{\mu} \psi^{-1}$, or for $\psi^{-1}$ and $\partial _{\mu} \psi$ and their contravariant counterparts. However, when we differentiate Eq.~(\ref{eq39}) with respect to the coordinates, we obtain
\begin{eqnarray}
\psi \,.\, \partial _{\mu} \psi^{-1} &=& - (\partial _{\mu} \psi) \, . \, \psi^{-1},
\nonumber \\
\psi^{-1} \, . \, \partial _{\mu} \psi &=& - (\partial _{\mu} \psi^{-1}) \, . \, \psi,
\label{eq40}
\end{eqnarray}
and identical formulas for the contravariant analogs.

Using Eqs.~(\ref{eq40}) and the property $\partial^{\mu}\partial_{\mu} \partial_{\nu} = \partial_{\nu}\partial^{\mu} \partial_{\mu}$, we obtain, after some calculations, the following identity
\be
\left(\psi^{-1} \partial^{\mu} \psi + \frac{1}{2} \partial^{\mu} \right) \left(\partial_{\mu} \psi^{-1} . \, \partial_{\nu} \psi + \psi^{-1} \partial_{\mu} \partial_{\nu}\psi \right) = \frac{1}{2} \, \psi^{-1} \, \partial_{\nu}(\partial^{\mu}\partial_{\mu}\psi \, . \, \psi^{-1}) \, . \, \psi,
\label{eq41}
\ee
which is the analog for (bi)quaternions of Eq.~(\ref{eq28}) for complex numbers.

Using this identity and replacing $\lambda_c$ by its value $\hbar / mc$ in Eq.~(\ref{eq38}), we obtain, after some rearrangements,
\be
\partial_{\nu} \left[ - \hbar^2(\partial^{\mu}\partial_{\mu} \psi \, . \, \psi^{-1}) - i \frac{\hbar e}{c}(A^{\mu} \partial_{\mu} \psi \, . \, \psi^{-1}) - i \frac{\hbar e}{c} \, \partial^{\mu} A_{\mu} + \frac{e^2}{c^2}A^{\mu} A_{\mu} \right] = 0,
\label{eq42}
\ee
which is a gradient that can be integrated. If we choose the integration constant to be $m^2 c^2$, we recover the Klein-Gordon equation in the form of Eq.~(\ref{eq31}), but now for a biquaternionic wave function.

\subsubsection{\it Spin-charge coupling term}
\label{scct}

In \cite{CN04,CN03}, we have shown how the free Dirac equation can be recovered from a free Klein-Gordon-like equation written in (bi)quaternionic form. Actually, a free (bi)quaternionic Klein-Gordon equation is equivalent to applying twice to the wave function the time part of the free Dirac operator while equating it to its spatial part.

This is no more the case when an exterior magnetic field enters into play. A new term  appears in this case and we show in the following that it corresponds to a coupling between the electron intrinsic magnetic moment and the magnetic field.

The standard Dirac equation for a spin 1/2 particle submitted to an exterior electromagnetic field can be written as
\be
\left(i \hbar \frac{\partial}{c \, \partial t} - \frac{e}{c}A^0 \right) \psi = \left[- \vec{\alpha} \left(i \hbar \vec{\nabla} + \frac{e}{c} \vec{A} \right) + mc \beta \right] \psi.
\label{eq43}
\ee

Using the Conway matrices (see e.g. Ref.~\cite{JS72}) corresponding to the Dirac matrices, we write the Dirac operator acting on $\psi$ as
\begin{eqnarray}
i \hbar \frac{\partial}{c \, \partial t} - \frac{e}{c}A^0 &=& {\rm e_3} \left(i \hbar \frac{\partial}{\partial x} + \frac{e}{c}A^x \right){\rm e_2} + {\rm e_1} \left(i \hbar \frac{\partial}{\partial y} + \frac{e}{c}A^y \right)i
\nonumber \\
&+& {\rm e_3} \left(i \hbar \frac{\partial}{\partial z} + \frac{e}{c}A^z \right){\rm e_1} + mc \, {\rm e_3}(\; \;){\rm e_3}.
\label{eq44}
\end{eqnarray}

We apply to $\psi$ the difference between the two squared parts of this Dirac operator, i.e. the square of the time part $TD$ corresponding to the l.h.s. of Eq.~(\ref{eq44}) and that of the spatial part $SD$ corresponding to its r.h.s., and we obtain, after some calculations,
\be
(TD)^2 - (SD)^2= KG + AT,
\label{eq45}
\ee
where $KG$ is the biquaternionic Klein-Gordon-like equation, written as
\begin{eqnarray}
\left(i \hbar \frac{\partial}{c \, \partial t} - \frac{e}{c}A^0 \right)^2 \psi &=& {\rm e_3}^2 \left(i \hbar \frac{\partial}{\partial x} + \frac{e}{c}A^x \right)^2 \psi \, {\rm e_2}^2 + {\rm e_1}^2 \left(i \hbar \frac{\partial}{\partial y} + \frac{e}{c}A^y \right)^2 \psi \, i^2
\nonumber \\
&+& {\rm e_3}^2 \left(i \hbar \frac{\partial}{\partial z} + \frac{e}{c}A^z \right)^2 \psi \, {\rm e_1}^2 + m^2c^2 \, {\rm e_3}^2(\psi) \, {\rm e_3}^2,
\label{eq46}
\end{eqnarray}
and $AT$ reads
\be
AT = \frac{\hbar e}{c}\left[{\rm e_2}(F_z^y \, . \, \psi){\rm e_1} + i (F_x^z \, . \, \psi){\rm e_3} - {\rm e_2}(F_y^x \, . \, \psi){\rm e_2}\right].
\label{eq47}
\ee

Since $F_z^y = -B^x$, $F_x^z = -B^y$ and $F_y^x = -B^z$, where $\vec{B}$ is the magnetic field, and since the Conway matrices involved correspond to minus the Dirac $\alpha$ matrices, $AT$ can be written as
\be
AT = \frac{\hbar e}{c} \, \vec{\alpha}. \vec{B} \, \psi.
\label{eq48}
\ee

This term expresses the coupling between the spin of the particle and the magnetic field. It is the relativistic analog of the term $-(e \hbar /2mc) \, \vec{\sigma} . \vec{B}$ involving the electron spin magnetic moment in the Pauli equation. Therefore, obviously, it cannot appear in the Klein-Gordon equation since this equation applies to a spin zero particle. However, the presence of this term in Eq.~(\ref{eq45}) prevents us from obtaining the standard form of the electromagnetic Dirac equation as a mere square root of a biquaternionic electromagnetic Klein-Gordon equation, as in the free case. We must therefore suspect that the electromagnetic Dirac equation proceeds from a different construction, which we detail in the following subsection.

\subsection{\bf \it Derivation of the electromagnetic Dirac equation}

The Dirac equation can be obtained by combining the quantum and scale covariances, but in a different way.

First we apply the quantum covariance as in \cite{CN04,CN03} and we obtain the following free Dirac equation,
\be
i \hbar \frac{\partial}{c \, \partial t} \, \psi = \left[- \vec{\alpha} \, i \hbar \vec{\nabla} + mc \beta \right] \psi.
\label{eq49}
\ee

Then we implement the scale covariance through the covariant derivative Eq.~(\ref{eq19}). Note that in Eq.~(\ref{eq19}) the total variation of the wave function is the sum of the inertial one (represented by the covariant derivative) and the geometric contribution $-i (e/ \hbar c)\,A_{\mu}$. This covariant derivative acts in a manner analogous to that encountered in general relativity where it amounts to subtracting the geometric contribution in order to keep only the inertial part. This is at variance with the way covariance is implemented by means of the quantum-covariant derivative, which includes nondifferentiability effects by adding new terms to the total derivative.

We therefore combine the quantum and scale covariances by replacing in the free Dirac equation (\ref{eq49}) the $\partial _{\mu }$ derivative by its inertial part ${\rm D}_{\mu }$ as given by Eq.~(\ref{eq19}). Thus, we obtain
\be
\left(i \hbar \frac{\partial}{c \, \partial t} - \frac{e}{c}A^0 \right) \psi = \left[- \vec{\alpha} \left(i \hbar \vec{\nabla} + \frac{e}{c} \vec{A} \right) + mc \beta \right] \psi.
\label{eq50}
\ee

Note that this is consistent with the way we have constructed the free equation as an inertial geodesic equation in a fractal space-time. We thus obtain the electromagnetic Dirac equation in its usual form as given by Eq.~(\ref{eq43}).

It must be stressed that such a method can also be employed to obtain the electromagnetic Klein-Gordon equation from the free one. This suggests that this way of combining quantum and scale covariances is grounded on more profound roots of the theory than the method used in Sec.~\ref{cekgge}.

Indeed, such a construction of the electromagnetic motion equations of quantum mechanics is formally analogous to the methods used in the standard field theory. However, while in this standard theory the form of the equations and of the QED covariant derivative are merely postulated as compatible with experiments, they have been given here more profound physical meanings grounded on the first principles of the scale relativity theory .


\section{\bf Conclusion}
\label{dc}


In the scale-relativistic framework, the electromagnetic Klein-Gordon equation had been previously obtained as a geodesic equation constructed as a strongly covariant combination of the quantum- and scale-covariant tools \cite{LN07}. Quantum covariance was implemented by the use of a complex covariant velocity operator and scale covariance by the use of a scale-covariant derivative which implemented the dilation/contraction transformations in the fractal space-time. The two tools were combined as in Eq.~(\ref{eq21}) to define a doubly (quantum and scale) covariant derivative which was used to write an inertial-like, strongly covariant geodesic equation. The electromagnetic Klein-Gordon equation proceeded from the integration of this equation. We have given a reminder of this derivation in Sec.~\ref{cekgge}.

Now, when we apply here the same procedure while using the biquaternionic 4-velocity obtained by taking into account the three levels of symmetry breaking implied by nondifferentiability, we actually obtain a biquaternionic electromagnetic Klein-Gordon-like equation. However, while in the free case the Dirac operator is the mere square root of the biquaternionic Klein-Gordon-like operator, this is no more the case when an electromagnetic field is involved. A new term appears which corresponds to a coupling of the spin of the electron to the magnetic field. The explanation is as follows.

In scale relativity, spin proceeds directly from the fractal geometry of space-time, while charges, which stem from transformations in scale space, are only indirect consequences of this fractality. Indeed, spin is an intrinsic quantum property of the particles, component of the total angular momentum, which is itself a constant of motion.

Here, the spin 1/2 can have two different interpretations. First it can be considered as a quantum charge of the electron/positron. In this case, in the same manner as Abelian and non-Abelian charges can be obtained in a geometric way from scale transformations in scale space \cite{NCL06}, spin arises geometrically from the fractality of space-time implied by both of the symmetry breakings ${\rm d}s\leftrightarrow - \, {\rm d}s$ and ${\rm d}x^{\mu}\leftrightarrow - \, {\rm d}x^{\mu}$ which are themselves consequences of non-differentiability \cite{CN06}. This interpretation corresponds to the way spin arises when the first strongly covariant method is used to obtain the motion equations. If we combine the two covariances (as done to recover the electromagnetic Klein-Gordon equation) when a nonzero spin particle is involved, a coupling term between the two charges, spin and electromagnetic, is mandatory.

Spin 1/2 can also be considered as directly linked to the fractal geometry of space-time. In this case, its coupling to the magnetic field is implicit and there is no additional term in the Dirac equation which recovers its usual form, as when the second method is used. This interpretation has been illustrated with numerical simulations aimed at visualizing the typically spinorial form of some geodesics in a fractal space-time (see Fig. 2 of Ref.~\cite{CN06}). Hence, the fractality influence through the spin is more fundamental and must be applied first, independently of that of the field.

This is the reason why the Dirac equation standard form can be obtained by combining the quantum and scale covariances in the following way. We first apply the quantum covariance as in Refs.~\cite{CN04} and \cite{CN03}. Then we implement the scale covariance through the QED covariant derivative previously recovered in the scale relativity framework. This allows us to obtain the electromagnetic Dirac equation under its usual form, which means that such a method encompasses more naturally the spin-charge coupling.

We have stressed that this scheme can also be employed to obtain the electromagnetic Klein-Gordon equation from the free one. This suggests that this way of combining quantum and scale covariances is grounded on more profound roots of the theory than merely combining them as it has been done in Ref.~\cite{LN07}.


\end{document}